# Exchange-spring behavior in bimagnetic CoFe$_2$O$_4$/CoFe$_2$ nanocomposite

Leite, G. C. P.[1], Chagas, E. F.[1], Pereira, R.[1], Prado, R. J.[1], Terezo, A. J.[2], Alzamora, M.[3], and Baggio-Saitovitch, E.[3]

[1] *Instituto de Física, Universidade Federal de Mato Grosso, 78060-900, Cuiabá-MT, Brazil*
[2] *Departamento de Química, Universidade Federal do Mato Grosso, 78060-900, Cuiabá-MT, Brazil*
[3] *Centro Brasileiro de Pesquisas Físicas, Rua Xavier Sigaud 150 Urca. Rio de Janeiro, Brazil.*

Phone number: 55 65 3615 8747

Fax: 55 65 3615 8730

Email address: efchagas@fisica.ufmt.br

## Abstract

In this work we report a study of the magnetic behavior of ferrimagnetic oxide CoFe$_2$O$_4$ and ferrimagnetic oxide/ferromagnetic metal CoFe$_2$O$_4$/CoFe$_2$ nanocomposites. The latter compound is a good system to study hard ferrimagnet/soft ferromagnet exchange coupling. Two steps were used to synthesize the bimagnetic CoFe$_2$O$_4$/CoFe$_2$ nanocomposites: (i) first preparation of CoFe$_2$O$_4$ nanoparticles using the a simple hydrothermal method and (ii) second reduction reaction of cobalt ferrite nanoparticles using activated charcoal in inert atmosphere and high temperature. The phase structures, particle sizes, morphology, and magnetic properties of CoFe$_2$O$_4$ nanoparticles have been investigated by X-Ray diffraction (XRD), Mossbauer spectroscopy (MS), transmission electron microscopy (TEM), and vibrating sample magnetometer (VSM) with applied field up to 3.0 kOe at room temperature and 50K. The mean diameter of CoFe$_2$O$_4$ particles is about 16 nm. Mossbauer spectra reveal two sites for Fe3+. One site is related to Fe in an octahedral coordination and the other one to the Fe3+ in a tetrahedral coordination, as expected for a spinel crystal structure of CoFe$_2$O$_4$. TEM measurements of nanocomposite show the formation of a thin shell of CoFe$_2$ on the cobalt ferrite and indicate that the nanoparticles increase to about 100 nm. The magnetization of nanocomposite showed hysteresis loop that is characteristic of the exchange spring systems. A maximum energy product (BH)$_{max}$ of 1.22 MGOe was achieved at room temperature for CoFe$_2$O$_4$/CoFe$_2$ nanocomposites, which is about 115% higher than the value obtained for CoFe$_2$O$_4$ precursor. The exchange-spring interaction and the enhancement of product (BH)$_{max}$ in nanocomposite CoFe$_2$O$_4$/CoFe$_2$ have been discussed.

# Introduction

The figure of merit for a permanent magnet material, the quantity $(BH)_{max}$ to the ideal hard material (rectangular hysteresis loop) is given by $(BH)_{max} = (2\pi M_S)^2$. For materials with high coercivity ($H_C$) the magnetic energy product is limited by the saturation magnetization ($M_S$). Aiming to overpass this limitation, and in order to obtain a material with high $(BH)_{max}$ product, Kneller and Hawig (1991) [1] proposed a nanocomposite formed by both hard (high $H_C$) and soft (high $M_S$) magnetic materials exchange coupled. These materials, called exchange spring or exchange-hardened magnets, combine the high coercivity of the hard material with the high saturation magnetization of the soft material, making possible the increase of the $(BH)_{max}$ product of the nanocomposite when compared with any individual phase that form the nanocomposite. [1-6].

The increase of the $M_S$ is caused by the exchange coupling between grains of nanometer size. Kneller and Hawig [1] derived a relationship that predicts how to reach a significant remanence enhancement using the microstructural and magnetic properties of this new kind of material, as the distribution of soft and hard magnetic phases and the fraction of soft magnetic phase, indicating the possibility of developing nanostructured permanent magnetic materials.

According to the exchange spring model of Kneller and Hawig, the critical dimension ($b_{cm}$) for the m-phase (soft material) depends on the magnetic coupling strength of the soft phase $A_m$ and the magnetic anisotropy of the hard phase $K_h$, according to the following equation:

$$b_{cm} = \pi \left(\frac{A_m}{2K_h}\right)^{1/2} \qquad \text{equation (1).}$$

To obtain a sufficiently strong exchange coupling, the grain size of the soft phase must be smaller than $2b_{cm}$. In a general way, a good magnetic coupling of the hard and soft components is achieved in materials with grain sizes of about 10–20 nm [7], the approximate value of the domain wall width in the hard magnetic materials.

Cobalt ferrite, $CoFe_2O_4$, is a hard ferrimagnetic material that has interesting properties like high $H_C$ [8, 9] moderate $M_S$ [10, 11], high chemical stability, wear resistance, electrical insulation and thermal chemical reduction [12, 13]. The latter property allows the transformation of $CoFe_2O_4$ in $CoFe_2$ (a soft ferromagnetic material with high $M_S$ value of about 230 $emu/g$ [14]) in moderate/high temperature. This property was used by Cabral *et. al.* [13] to obtain the nanocomposite $CoFe_2O_4/CoFe_2$ and by Scheffe *et al.* to hydrogen production [12]. Also, the $CoFe_2O_4/CoFe_2$ nanostrutured bimagnetic material was formerly studied as layered thin films by Jurca *et. al.* [15] and Viart *et. al.* [16].

In this work we describe an original process of chemical reduction used for the synthesis of the $CoFe_2O_4/CoFe_2$ nanocomposite materials, as well as the magnetic and structural characterization of both precursor and nanocomposite materials. Finally, the enhancement obtained for the $(BH)_{max}$ product of the $CoFe_2O_4/CoFe_2$ nanocomposite compared with that of the $CoFe_2O_4$ precursor is reported.

## Experimental procedure

### Synthesis of $CoFe_2O_4$

The hydrothermal method was used to synthesize cobalt ferrite. This method provides different classes of nanostructurated inorganic materials from aqueous solutions, by means of small Teflon autoclaves and has a lot of benefits such as: clean product with high degree of crystallinity at a relative low reaction temperature (up to 200ºC). All the reagents used in this synthesis are commercially available and were used as received without further purification. An appropriate amount of analytical-grade ammonium ferrous sulfate $((NH_4)_2(Fe)(SO_4)_2 \cdot 6H_2O$ (0.5 g, 1.28 mmol) and sodium citrate $Na_3C_6H_5O_7$ (0.86 g, 4.72 mmol) was dissolved in 20 ml of ultra pure water and stirred together for 30 min at room temperature, then stoichiometric $CoCl_2 \cdot 6H_2O$ (0.15 g, 0.64 mmol) was added and dissolved, followed by the addition of an aqueous solution of 5M NaOH. The molar ratio of Co (II) to Fe (II) in the above system was 1:2. The mixtures were transferred into an autoclave, maintained at 120 °C for 24 h and then cooled to room temperature naturally. A blackish precipitate was separated and several times washed with ultra pure water and ethanol.

### Synthesis *$CoFe_2O_4$/$CoFe_2$* Nanocomposite

To obtain the nanocomposite we mixed the nanoparticles of cobalt ferrite with activated charcoal (carbon) and subjected the mixture to heat treatment at 900 °C for 3 hours in inert atmosphere (Ar), promoting the following chemical reduction:

$$CoFe_2O_4 + 2C \xrightarrow{\Delta} CoFe_2 + 2CO_2$$

The symbol Δ indicates that thermal energy is necessary in the process.
The similar process was used by Cabral *et. al*. [13] to obtain the same nanocomposite and by Scheffe *et. al*. to produce hydrogen[12].
Theoretically, varying the molar ratio between activated carbon and cobalt ferrite we can control the formation of $CoFe_2$ phase in the nanocomposite. However, the process is difficult to control due the residual oxygen in the inert atmosphere.
Two samples were prepared using the process described here: a full and another partially reduced. The molar ratio between activated charcoal and cobalt ferrite was 2:1 and 10:1, to the partially and fully reduced samples respectively.

### Structural and magnetic measurements

The crystalline phases of the calcined particles were identified by the powder X-ray diffraction (XRD) patterns of the magnetic nanoparticles were obtained on a Siemens D5005 X-ray diffractometer using Cu-K radiation (0.154178 nm).
Magnetic measurements were carried out using a VSM (VersaLab Quantum Design) at room temperature and 50K. $^{57}Fe$ Mossbauer spectroscopy experiments were performed in two temperatures, 4.2 and 300 K to $CoFe_2O_4$ samples.

The morphology and particle size distribution of the samples were examined by direct observation via transmission electron microscopy (TEM) (model JEOL-2100, Japan).

## Results and Discussion

The XRD analysis of the synthesized powder after calcination (figure 1) shows that the final product is $CoFe_2O_4$ with the expected inverse spinel structure (JCPDS No. 00-022-1086), presenting the Fd3m spatial group with a lattice parameter $a$ = 8.403Å ± 0.0082 Å. Value close to that is expected for the bulk $CoFe_2O_4$ ($a$ = 8.39570) [17]. The XRD pattern also reveals traces of Co and $Co_7Fe_3$ crystalline phases (indicated in figure 1).

Figure 2 shows the diffraction profile obtained for the sample completely reduced. The XRD profile is similar to that to the $CoFe_2$ (JCPDS No. 03-065-4131), indicating the expected chemical reduction occurred. Due the small quantity of the sample partially reduced obtained we could not perform XRD measurements.

To analyze the cation distribution of the precursor compound ($CoFe_2O_4$), Mossbauer spectroscopy experiments at room temperature and 4.2 K were performed, as shown in the figure 3. The Mossbauer measurements at 4.2 K reveals two sites for $Fe^{3+}$ related to both octahedral and tetrahedral coordination, respectively, as expected for the spinel crystal structure of $CoFe_2O_4$ [18].

The morphology and dimension of nanoparticles were analyzed by TEM measurements. The measurement of the cobalt ferrite sample (precursor material) shows formation of aggregates. This result is expected to samples prepared by hydrothermal method [19, 20]. Figure 4 shows a TEM image of cobalt ferrite particles. The TEM measurement reveals that the $CoFe_2O_4$ nanoparticles form a polidisperse system with approximately spherical nanoparticles. The of particle size distribution indicates that ferrite cobalt particles have mean diameter of 16 nm and the standard deviation of about 4.9 nm. The particle size histogram obtained by TEM measurements of the cobalt ferrite sample is shown in figure 5.

The TEM measurements of the nanocomposite ($CoFe_2O_4/CoFe_2$) are shown in figures 6 and 7. In figure 6a one can see there is roughness at the surface of the nanoparticle. Note that similar roughness was not observed at the surface of the precursor material (figure 4). In addition, figure 6b shows that the superficial material connects the nanoparticles and the most part of this material is in the interface of the nanoparticles. In figure 7a one can see that the nanoparticle is composed of two parts a big core and a thin shell (thickness about 1.5 nm). Similar pictures are observed to other nanoparticles (not shown). As previously mentioned, the shell does not cover each nanoparticle but the aggregates of nanoparticles. We attribute the core to the $CoFe_2O_4$ (hard material) and the shell to the $CoFe_2$ (soft material). Thus the nanocomposite obtained is constituted of spheres of magnetically hard material in a soft matrix.

The inserts in figures 7a and 7b show details of the interplanar distance of both core and shell, respectively. The interplanar distance observed to the core is about 0.49 nm (insert of figure 7a). This value is the same obtained by Chen *et. al* [21] to the (111) plane of $CoFe_2O_4$. The insert in figure 7b shows an interplanar distance of about 0.3 nm, obtained to the shell. But due the small thickness of the shell we consider necessary measurements of high-resolution TEM (HRTEM) to more precise results.

TEM analysis indicates that the nanoparticles of nanocomposite are larger than the originals nanoparticles, indicating the reduction process increases the mean size (diameter) of the nanoparticles to about 100 nm. Also, TEM measurements showed that the dimension of the soft phase ($CoFe_2$) is larger than the critical size obtained by equation 1 (see figure 6b). Using the magnetic parameters available for $CoFe_2O_4$ and $CoFe_2$ ($A_m \sim 1.7 \times 10^{-11}$ J/m [22, 23], $K_h \sim 2.23 \times 10^5$ J/m$^3$)[24], the calculated critical grain size $b_{cm}$ for the soft $CoFe_2$ phase is about 20 nm.

The cobalt ferrite sample studied in this work has shown coercivity about 1.69 kOe, at room temperature. This result is higher than the coercivity obtained by Cabral *et. al.* (1.32 kOe) [13] but lower than those reported by Ding *et. al.* [8] and Liu *et. al.* [9] to samples treated by thermal magnetic annealing and mechanical milling, respectively.

The hysteresis loop at 50K shows a strong increase of coercivity (8.8 kOe) compared with the value obtained at room temperature (see the figure 8). Similar behavior of coercivity was observed by Maaz *et. al.* [25] and Gopalan *et. al.* [26]. Another effect observed by theses authors and also observed in this work is the increase of the remanence ratio ($M_r/M_S$). The saturation magnetization ($M_S$) and remanent magnetization ($M_r$) obtained here were, respectively, 445 emu/cm$^3$ (82 emu/g) and 181 emu/cm$^3$ (33 emu/g) at room temperature, while at 50 K were 477 and 323 emu/cm$^3$ (88 and 60 emu/g). These values indicate an increase for remanence ratio ($M_r/M_S$), from 0.42 to 0.68, when the temperature is decreased from 300 K to 50 K. In these results there are two important facts: first the increase of $M_r/M_S$ value; and second, the $M_r/M_S$ value obtained at room temperature is close to the theoretical value expected (0.5) to non interacting single domain particles with uniaxial anisotropy [27] even the cobalt ferrite has a cubic structure. Kodama [28] attribute the existence of an effective uniaxial anisotropy in magnetic nanoparticles to the surface effect. The strong anisotropy that produces a high coercivity can also caused by surface effect [28]. Golapan *et. al.* [26] suggest that the increase in the value of the $M_r/M_S$ ratio is associated with an enhanced of cubic anisotropy contribution at lower temperature.

Figure 9 shows the hysteresis loop of the sample partially reduced ($CoFe_2O_4/CoFe_2$) at room temperature and 50K. The hysteresis curves of the nanocomposite can be described by a single-shaped loop (no steps in the loop) similar to that of a single phase indicating that magnetization of both phases reverses cooperatively.

The same behavior observed to coercivity for the cobalt ferrite was also seen for the nanocomposite. The coercivity increased from 1.34 kOe (at 300K) to 6.0 kOe (at 50K). This enormous increase of coercivity deserves more investigation.

The $M_S$ obtained at room temperature was about 146 *emu/g*, a value is higher than the $M_S$ obtained for precursor material and lower than the expected for pure $CoFe_2$ (230 *emu/g*) [14].

The $CoFe_2O_4/CoFe_2$ nanocomposites demonstrate inter-phase exchange coupling between the magnetic hard phase and the magnetic soft phase, which lead to magnets with improved energy products. We obtained an energy product $(BH)_{max}$ of 1.22 MGOe to the nanocomposite. This value is about 115% higher than the value obtained for $CoFe_2O_4$. To room temperature we obtained 0.568 MGOe to the product $(BH)_{max}$, assuming the theoretical density for $CoFe_2O_4$ [29]. The value of $(BH)_{max}$ to the nanocomposite is higher than the best value obtained by Cabral *et. al.* (0.63 MGOe) to same nanocomposite (but with different molar ratio). Also, the precursor sample prepared in this work showed higher coercivity and saturation magnetization than the precursor sample of Cabral *et. al.*. This observation suggest that the $(BH)_{max}$ product depends of the magnetic properties of precursor material.

Considering the nanocomposite formed only by the mixture of $CoFe_2O_4$ and $CoFe_2$, we expect that the value of $M_S$ is the sum of individual saturation magnetization of these two compounds. Using the values of $M_S$ = 230 *emu/g* to $CoFe_2$ [14] and 82 *emu/g* to cobalt ferrite (result of this work), the saturation magnetization of the nanocomposite (146 *emu/g*) suggests that content of $CoFe_2$ in the nanocomposite is about 40% and that of $CoFe_2O_4$ 60 %.

To better visualization the improvement obtained in magnetic properties, figure 10 show both of hysteresis curve of the nanocomposite $CoFe_2O_4/CoFe_2$ and cobalt ferrite at room temperature. The small decrease of coercivity and the increase of $M_r$ and $M_S$ are expected. These behaviors can be qualitatively explained by the simple one-dimensional model proposed by Kneller and Hawig.

## Conclusion

We synthesize nanocomposite of hard ferrimagnetic $CoFe_2O_4$ and soft ferromagnetic $CoFe_2$ with exchange spring behavior at room temperature. This assertion is confirmed by the hysteresis curve of the nanocomposite, which do not show steps in the loop. The thermal treatment at 900 °C used in the synthesis method increases the mean size of nanoparticles to about 100 nm (indicated by TEM measurements). However the thermogravimetric analysis (not shown) indicates that the similar treatment can be used at temperature about 600°C, increasing the time.

The chemical reduction process described in this work is a good pathway to obtain $CoFe_2O_4/CoFe_2$ with exchange spring behavior, but the residual oxygen in the argon commercial gas makes difficult the control the $CoFe_2$ molar ratio in the nanocomposite.

The magnetic energy product was greatly improved in the nanocomposite when compared with the ferrite precursor. However, other studies show that the coercivity of this precursor material can be increased by thermal annealing, thermal and magnetic annealing or mechanical milling. This increase of coercivity may also improve the magnetic energy product of the nanocomposite, but this assumption deserves more investigation.

## Acknowledgments

This work has been supported by Brazilian funding agency CAPES (PROCAD-NF 2233/2008). The authors would like to thank the LME/LNLS for technical support during electron microscopy work.

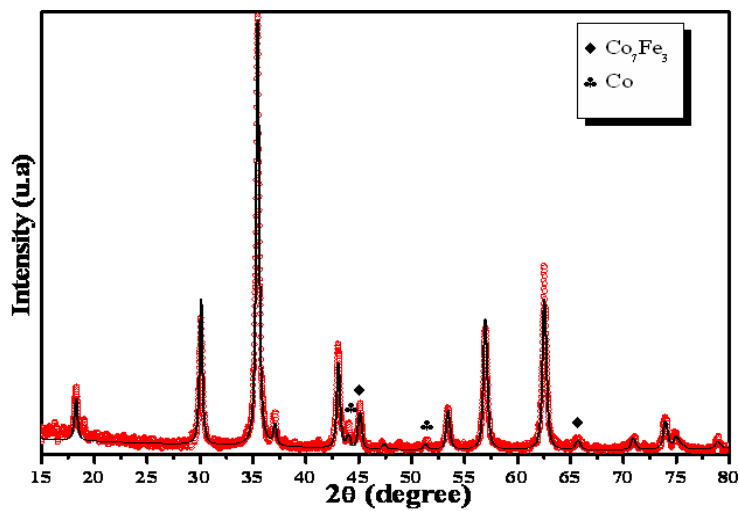

Figure 1 – XRD diffraction patterns of the $CoFe_2O_4$. Rietveld fits (solid line) are displayed.

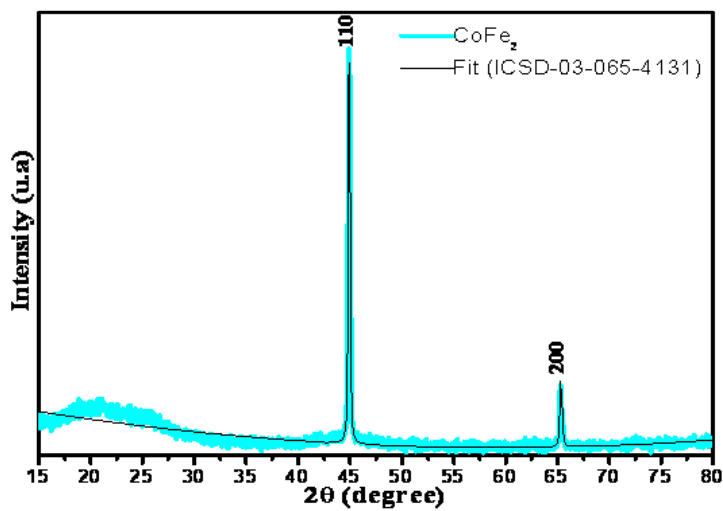

Figure 2 - XRD diffraction patterns of the $CoFe_2$ produced by reduction reaction of cobalt ferrite nanoparticles blended with activated charcoal in the molar ratio 1:10.

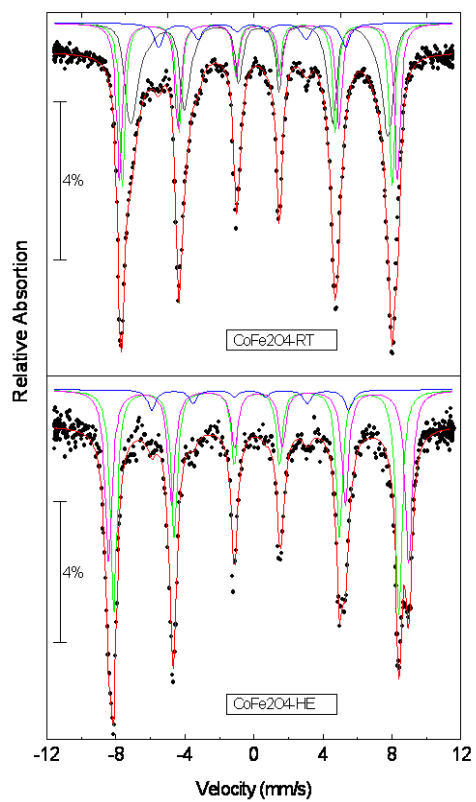

Figure 3 – Mossbauer spectra of $CoFe_2O_4$ at room temperature (RT) and 4,2 K (He).

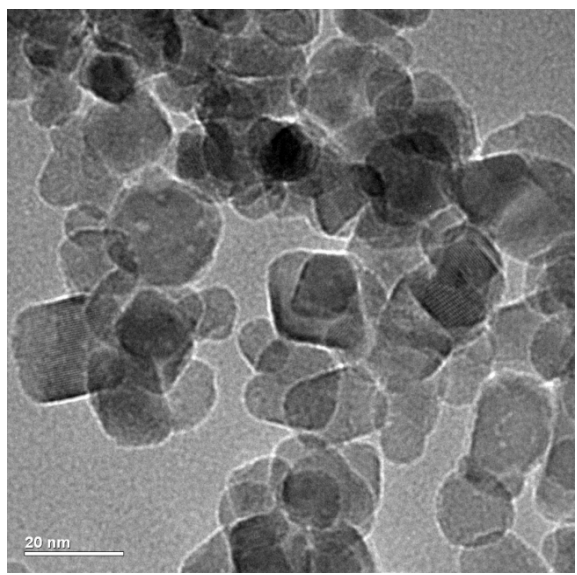

Figure 4 - Transmission electron microscopy of as-prepared $CoFe_2O_4$ by hydrothermal method.

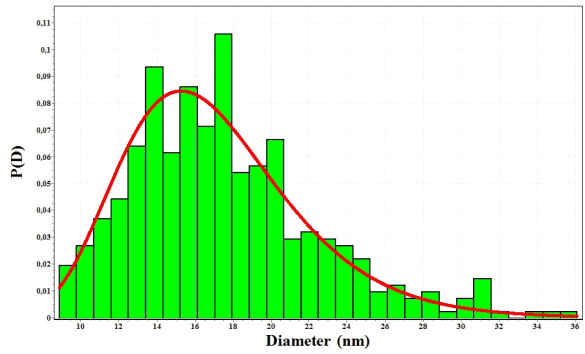

Figure 5– Histogram of the particle size distribution, fitted by a log normal distribution (solid line). Particles have mean diameters of 16 nm.

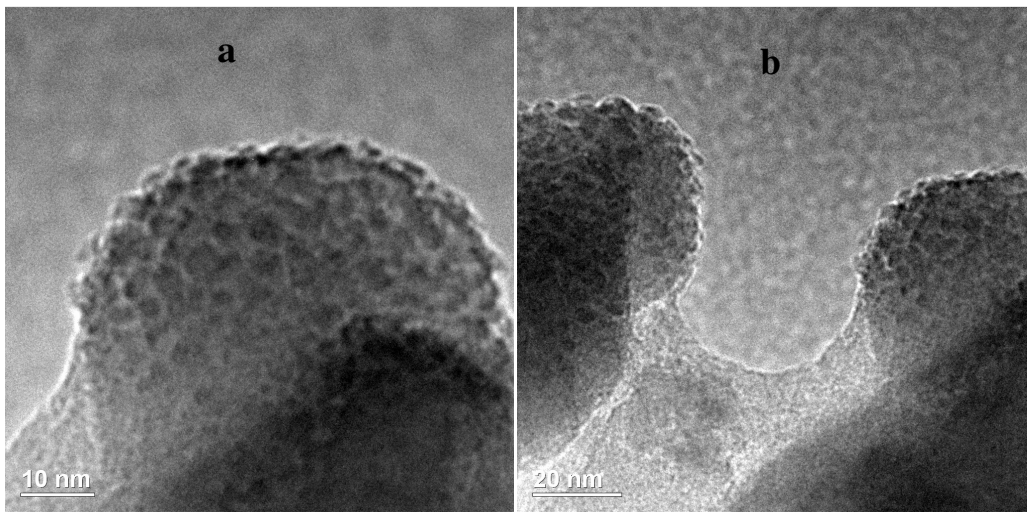

Figure 6 – Transmission electron microscopy of nanocomposite $CoFe_2O_4/CoFe_2$. View of a) one nanoparticle and b) two nanoparticles.

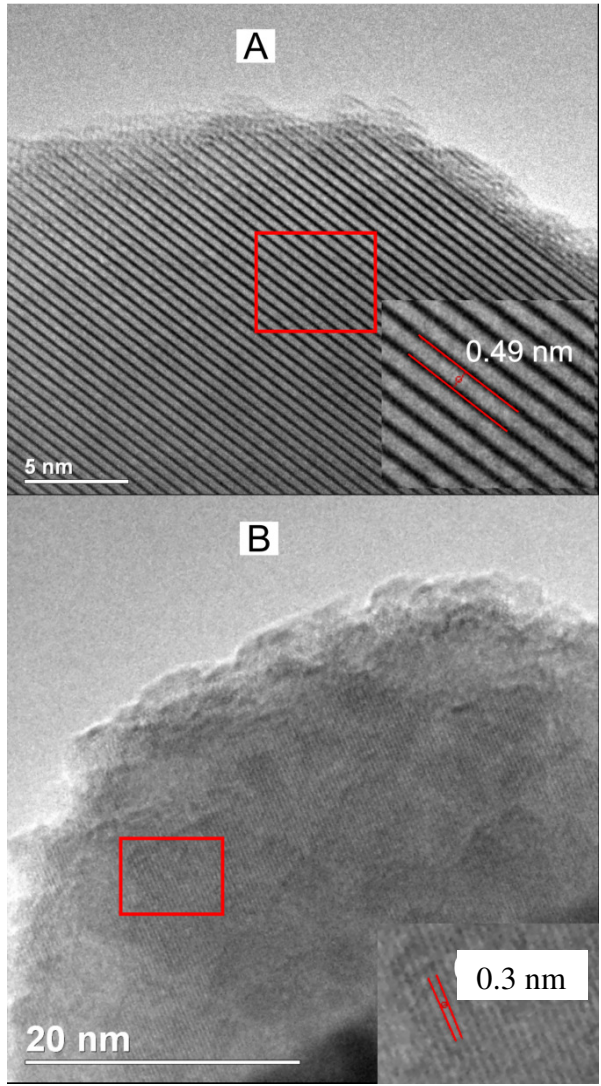

Figure 7– TEM measurements of nanocomposite $CoFe_2O_4/CoFe_2$. (A) Show the an interplanar distance of $CoFe_2O_4$ the insert show details of marked area. (B) Show the an interplanar distance of $CoFe_2$ the insert show details of marked area.

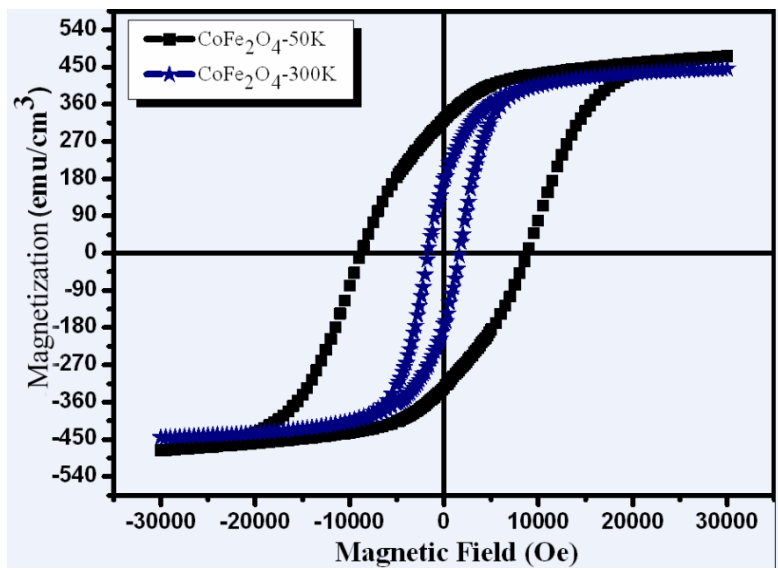

Figure 8 – Hysteresis loops $CoFe_2O_4$ nanoparticles at room temperature (300 K) and 50 K at maximum applied field of 30kOe.

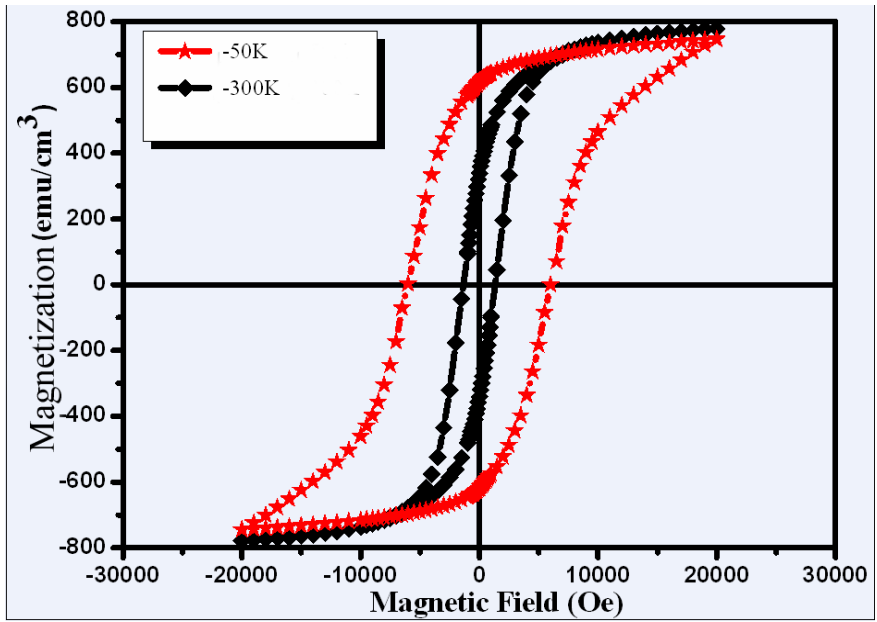

Figure 9 - Hysteresis loops for $CoFe_2O_4/CoFe_2$ nanocomposites at room temperature (300 K) and 50 K at maximum applied field of 20 kOe

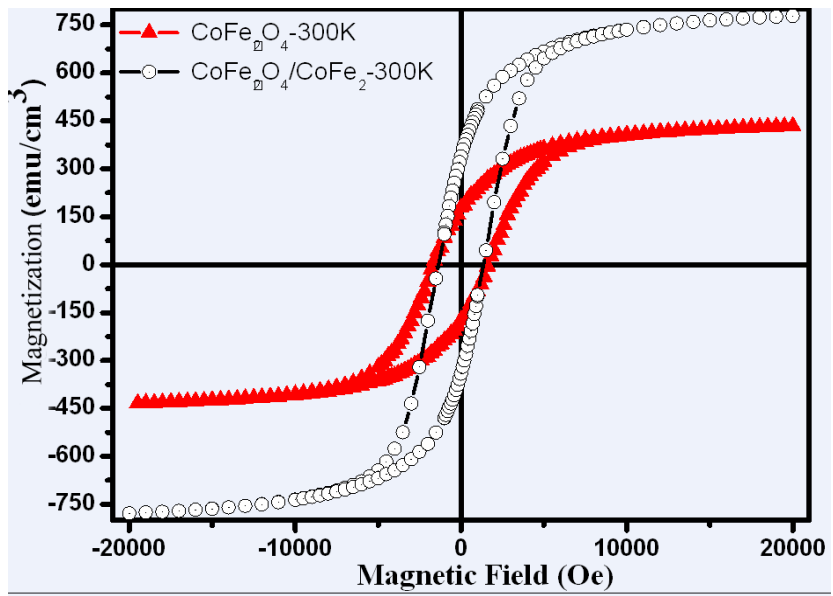

Figure 10 – Hysteresis loops to $CoFe_2O_4/CoFe_2$ and $CoFe_2O_4$ nanocomposites both at room temperature (300 K).

Reference

[1] E.F. Kneller, R. Hawig, The Exchange-Spring Magnet: A New Material Principle for Permanent Magnets, Journal of Magnetism and Magnetic Materials, 27 (1991).
[2] L. Withanawasam, A.S. Murphy, G.C. Hadjipanayis, R.F. Krause, Nanocomposite R2Fe14B/Fe exchange coupled magnets, J. Appl. Phys, 76 (1994) 7065-7067.

[3] T. Schrefl, J. Fidler, H. Kronmüller, Remanence and coercivity in isotropic nanocrystalline permanent magnets, Phys. Rev. B, 49 (1994) 6100-6110.

[4] R. Skomski, J.M.D. Coey, Giant energy product in nanostructured two phase magnets, Phys. Rev. B, 48 (1993) 15812-15816.

[5] H. Zeng, S. Sun, J. Li, Z.L. Wang, J.P. Liu, Tailoring magnetic properties of core/shell nanoparticles, Appl. Phys. Lett., 85 (2004) 792-794.

[6] E.E. Fullerton, J.S. Jiang, S.D. Bader, Hard/soft magnetic heterostructures: model exchange-spring magnets, Journal of Magnetism and Magnetic Materials, 200 (1999) 392-404.

[7] A. Hernando, J.M. González, Soft and hard nanostructured magnetic materials, Hyperfine Interactions, 130 (2000) 221-240.

[8] J. Ding, P.C. McCormick, R. Street, Magnetic Properties of Mechanically Alloyed CoFe2O4, Solid State Communications, 95 (1995) 31-33.

[9] B.H. Liu, J. Ding, Strain-induced high coercivity in CoFe2O4 powders, Appl. Phys. Lett., 88 (2006) 042506-042508.

[10] T.L. Templeton, A.S. Arrott, A.E. Curzon, M.A. Gee, X.Z. Li, Y. Yoshida, P.J. Schurer, J.L. LaCombe, Magnetic properties of CoxFe3−xO4 during conversion from normal to inverse spinel particles, J. Appl. Phys, 73 (1993) 6728-6730.

[11] M. Rajendran, R.C. Pullar, A.K. Bhattacharya, D. Das, S.N. Chintalapudi, C.K. Majumdar, Magnetic properties of nanocrystalline CoFe2O4 powders prepared at room temperature: variation with crystallite size, J. Magn. Magn. Mater., 232 (2001) 71-83.

[12] J.R. Scheffe, M.D. Allendorf, E.N. Coker, B.W. Jacobs, A.H. McDaniel, A.W. Weimer, Hydrogen Production via Chemical Looping Redox Cycles Using Atomic Layer Deposition-Synthesized Iron Oxide and Cobalt Ferrites, Chem. Mater. , 23 (2011) 2030-2038.

[13] F.A.O. Cabral, F.L.A. Machado, J.H. Araújo, J.M. Soares, A.R. Rodrigues, A. Araújo, Preparation and Magnetic Study of the CoFe2O4 – CoFe2 Nanocomposite Powders, IEEE Transactions on Magnetics, 44 (2008) 4235-4238.

[14] M. Mohan, V. Chandra, S.S. Manoharan, A New Nano CoFe2 Alloy Precursor for Cobalt Ferrite Prodution Via Sonoreduction Process, Current Science, 94 (2008) 473-476.

[15] I.S. Jurca, N. Viart, C. Mény, C. Ulhaq-Bouillet, P. Panissod, G. Pourroy, Structural Study of CoFe2O4/CoFe2 Multilayers, Surface Science, 529 (2004) 215-222.

[16] N. Viart, R.S. Hassan, J.L. Loison, G. Versini, F. Huber, P. Panissod, C. Mény, G.d. Pourroy, Exchange Coupling in CoFe2O4/CoFe2 Bilayers Elaborated by Pulsed Laser Deposition, Journal of Magnetism and Magnetic Materials, 279 (2004) 21-26.

[17] G. Bate, E.B. Wohlfarth, Ferromagnetic Materials, North-Holland, Amsterdam, 1980.

[18] H.H. Hamdeh, W.M. Hikal, S.M. Taher, J.C. Ho, N.P. Thuy, O.K. Quy, N. Hanh, Mössbauer evaluation of cobalt ferrite nanoparticles synthesized by forced hydrolysis, Journal of Applied Physics, 97 (2005) 064310-064310-064304.

[19] M.G. Naseri, E.B. Saion, H.A. Ahangar, A.H. Shaari, M. Hashim, Simple Synthesis and Characterization of Cobalt Ferrite Nanoparticles by a Thermal Treatment Method, J. Nanomat., 2010 (2010 ) 1-8.

[20] S.C. Goh, C.H. Chia, S. Zakaria, M. Yusoff, C.Y. Haw, S. Ahmadi, N.M. Huang, H.N. Lim, Hydrothermal preparation of high saturation magnetization and coercivity cobalt ferrite nanocrystals without subsequent calcination, Mater. Chem. and Phys., 120 (2010) 31-35.

[21] Z. Chen, L. Gao, Synthesis and magnetic properties of CoFe2O4 nanoparticles by using PEG as surfactant additive, Material Science & Enginnering B, 141 (2007) 82-86.

[22] X. Liu, A. Morisako, Soft magnetic properties of FeCo films with high saturation magnetization, J. Appl. Phys, 103 (2008) 07E726-707E728.

[23] M.A. Willard, D.E. Laughlin, M.E. McHenry, Recent advances in the development of (Fe,Co)88M7B4Cu1 magnets, J. Appl. Phys. , 87 (2000) 7091-7096.

[24] A.J. Rondione, A.C.S. Samia, Z.J. Zhang, Characterizing the magnetic anisotropy constant of spinel cobalt ferrite nanoparticles, Appl. Phys. Lett., 76 (2000).

[25] K. Maaz, A. Mumtaz, S.K. Hasanain, A. Ceylan, Synthesis and magnetic properties of cobalt ferrite (CoFe2O4) nanoparticles prepared by wet chemical route, Journal of Magnetism and Magnetic Materials, 308 (2007) 289-295.

[26] E.V. Gopalan, I.A. Al-Omari, D.S. Kumar, Y. Yoshida, P.A. Joy, M.R. Anantharaman, Inverse magnetocaloric effect in sol–gel derived nanosized cobalt ferrite, Applied Phisics A, 99 (2010) 497-503.

[27] T. Ibusuki, S. Kojima, O. Kitakami, Y. Shimada, Magnetic Anisotropy and Behaviors of Fe Nanoparticles, IEEE Transactions on Magnetics, 37 (2001) 2223-2225.

[28] R.H. Kodama, Magnetic nanoparticles, Journal of Magnetism and Magnetic Materials, 200 (1999) 359-372.

[29] A. Rafferty, T. Prescott, D. Brabazon, Sintering behavior of cobalt ferrite ceramic, Ceram Int, 34 (2008) 15-21.